\documentclass[conference]{IEEEtran}

\usepackage{myStyleIEEE}

\IEEEoverridecommandlockouts

\makeatletter
\def\ps@IEEEtitlepagestyle{%
  \def\@oddfoot{\mycopyrightnotice}%
  \def\@evenfoot{}%
}
\def\mycopyrightnotice{%
  {978-1-5386-7138-2/18/\$31.00~\copyright~2018 IEEE \hfill}
  \gdef\mycopyrightnotice{}
}

\begin{document}

\title{An Equivalent Circuit Formulation for Power System State Estimation including PMUs \vspace*{-2mm}}

\renewcommand{\theenumi}{\alph{enumi}}
\renewcommand{\arraystretch}{1.3}
\author{
\IEEEauthorblockN{Aleksandar Jovicic\IEEEauthorrefmark{1}, Marko Jereminov\IEEEauthorrefmark{2}, Larry Pileggi\IEEEauthorrefmark{2}, Gabriela Hug\IEEEauthorrefmark{1}}%
\IEEEauthorblockA{\IEEEauthorrefmark{1} EEH - Power Systems Laboratory, ETH Zurich, Zurich, Switzerland } %
\IEEEauthorblockA{\IEEEauthorrefmark{2} Department of Electrical and Computer Engineering, Carnegie Mellon University, Pittsburgh, PA\\
Emails: \{jovicic, hug\}@eeh.ee.ethz.ch, \{mjeremin, pileggi\}@andrew.cmu.edu \vspace*{-0.525cm}}

}

\maketitle
\IEEEpeerreviewmaketitle


\begin{abstract}
In this paper, a novel formulation for the power system state estimation is proposed, based on the recently introduced equivalent split-circuit formulation of the power flow problem. The formulation models the conventional and time synchronized measurements simultaneously and contains a significantly lower level of nonlinearity compared to the available hybrid state estimators. The appropriate circuit models are derived for different types of measurements and integrated into the existing circuit framework for the power flow problem. A constrained optimization problem is then formulated to estimate the states of the system in rectangular coordinates, while satisfying the circuit equations and bounds on the measurement data. To further prove the concept and validate the accuracy of the proposed formulation, several test cases are solved and the results are presented and discussed. \\
\end{abstract}


\section{Introduction}
State estimation (SE) is a vital part of modern power network monitoring systems. It processes raw noisy measurements from meters installed in the grid and estimates the system operating state. This estimate is normally given as a set of voltage phasors for the buses in the system. Based on the SE output, system operators can confidently execute different network analyses and respond appropriately. Since F. Schweppe established the fundamentals of SE in the 1970s \cite{Schweppe}, researchers have worked to refine this tool and improve computational efficiency, robustness, and accuracy.

For several decades, power system measurements have been obtained by Remote Terminal Units (RTU). These devices typically provide information about voltage magnitude, as well as active and reactive power flows and injections. Therefore, SE algorithms were initially based solely on these data. To this day, the nonlinear Weighted Least Square (WLS) method remains the standard approach used in conventional SE \cite{Abur}. The high degree of nonlinearity, intrinsic to WLS, is mainly a consequence of the trigonometric relations between active and reactive power flows and injections, and voltage angles.  
                    
Due to increased penetration of Phasor Measurement Units (PMU), the SE area is undergoing significant changes, as deployment of PMUs provide highly accurate, time synchronous, measurements of voltage and current phasors, thus not only improving the estimation, but also rendering the SE problem linear in case the system is entirely observable by PMUs \cite{Phadke1985}. However, this scenario is highly unlikely in the near future, considering the still relatively high costs of PMU devices. Thus, the main challenge remains how to most effectively combine PMU and RTU measurements, as both provide different but important information about the state of the system. Some of the key issues are the treatment of the current phasor measurements, as well as the significant difference in accuracy of PMU and RTU data. 

Numerous hybrid estimators have been proposed in literature. In \cite{Ming2006}, a sequential hybrid state estimator is presented. The states are estimated at two separate stages; first, only conventional RTU measurements are considered, while the second stage comprises only PMU synchronized measurements and is therefore linear. Standard SE algorithms are deployed in both steps and the different types of measurements are observed separately. However, the second stage estimator utilizes estimation results from the first stage as an initial point, which is a drawback of this method. Multi-stage estimators based on separate processing of conventional and synchrophasor measurements are presented also in \cite{Baltensperger2010,Costa2013,Wu2017}. All of these methods assume full system observability by the RTU measurements. 

The approaches proposed in \cite{Bi2008} and \cite{Chakrabarti2010} combine both PMU and RTU measurements into a single-stage estimation problem, with bus voltage magnitudes and angles as the unknown states. The method in \cite{Bi2008} utilizes the PMU current measurements in rectangular form and the appropriate relations between those measurements and bus voltages in polar form are derived. In \cite{Chakrabarti2010} on the other hand, the PMU current measurements are used in polar coordinates, but the same as in \cite{Bi2008}, are still expressed as trigonometric functions of the bus voltage angles. A modification of the state vector to include current phasor measurements directly in polar form is discussed in \cite{Valverde2011}, thus avoiding any measurement transformations. However, in all of these approaches, standard SE is then applied to the resulting measurement functions. 

Recent advances in the steady-state analysis of power systems introduced the equivalent split-circuit formulation for simulating the power flow problem \cite{CMU1,CMU2,CMU3,CMU4}. In these works, an equivalent circuit representation of the power system consisting of coupled circuits has been derived in terms of the current and voltage state variables in rectangular form. It was shown that the equivalent circuit formulation allows for a comprehensive application of circuit simulation methods to a range of large scale power system problems, hence guaranteeing the stable and robust convergence of these problems. More importantly, the circuit nature of the proposed formulation allows for trivially handling the current and voltage phasor measurements, that can be directly incorporated within the equivalent circuit framework.

In this paper, we extend the equivalent split-circuit formulation to formulate the novel equivalent circuit based static state estimator. Unlike the existing SE algorithms, the proposed formulation does not rely on standard measurement functions incorporated into a WLS or Least Absolute Value algorithm. Instead, circuit models for various types of conventional and time synchronized measurements are derived and incorporated within the equivalent circuit framework that represents the power flow problem\cite{CMU1,CMU2,CMU3,CMU4}. Subsequently, an optimization problem is formulated based on the new equations of the system equivalent circuit and using interval arithmetic to translate RTU measurements into circuit components. System states are then estimated in rectangular form by solving this optimization problem. In this way, RTU and PMU measurements are treated simultaneously, while the degree of nonlinearity is significantly reduced compared to the previously mentioned algorithms.   

The remainder of the paper is structured as follows. Sect. \ref{sec: 2} gives a brief introduction into the equivalent circuit formulation of the power flow problem, as well as basic principles of interval arithmetic. In Sect. \ref{sec: 3}, circuit models for different types of measurements are derived. Further, in Sect. \ref{sec: 4} the SE optimization problem is formulated. The proposed approach is implemented on several test systems and results are presented in Sect. \ref{sec: 5}. Finally, Sect. \ref{sec: 6} concludes the paper. 

\section{Background} \label{sec: 2}
\subsection{Equivalent Circuit Formulation}
An equivalent circuit formulation (ECF) of the power flow problem was introduced in \cite{CMU1,CMU2,CMU3,CMU4}. The main idea of this approach is that each separate element of the power system (e.g. generator, load, transmission line etc.) can be represented by an equivalent circuit model based on the relationship between current and voltage at its terminals. Therefore, any physics based model can easily be incorporated into this formulation. The nonlinear equivalent circuit models are generally handled by using their linearized models obtained from the first order Taylor approximation and iteratively solving the linearized circuit until convergence, which is equivalent to Newton-Raphson (N-R) iterations. However, the circuit nonlinearities in the power flow problem are defined by a complex conjugate operator, which is not analytic and as such prevents the use of first derivative and Taylor expansion. Therefore, one of the core steps toward successful implementation of ECF is splitting the power system equivalent circuit into two sub-circuits, coupled by controlled sources \cite{CMU1}. The first sub-circuit includes the real part of voltages and currents, while the second sub-circuit includes the imaginary parts. As a result, the underlying system of circuit equations contains only real values and the N-R iterations can be applied (see \cite{CMU1,CMU2,CMU3,CMU4} for more details).

\begin{figure}[h!]
    \vspace{-0.175cm}
    \centering
    \begin{subfigure}[h]{0.135\textwidth}
        \centering  
        \includegraphics[width=\textwidth]{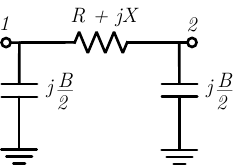}
        \caption{\hspace*{0.25em}}  
        \label{fig:1a}
    \end{subfigure}
    \hfill
    \begin{subfigure}[h]{0.165\textwidth}  
        \centering 
        \includegraphics[width=\textwidth]{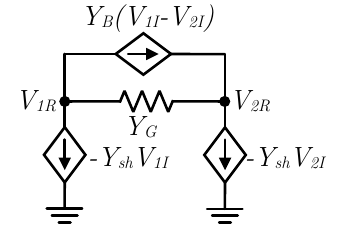}
        \caption{\hspace*{0.25em}}  
        \label{fig:1b}
    \end{subfigure}
    \hfill
        \begin{subfigure}[h]{0.165\textwidth}  
        \centering 
        \includegraphics[width=\textwidth]{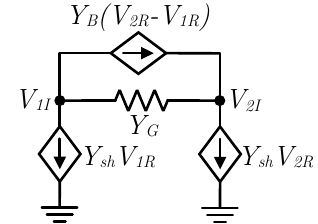}
        \caption{\hspace*{0.25em}}  
        \label{fig:1c}
    \end{subfigure}
    \caption{\label{fig:1}(a) Pi-model for a transmission line; (b) steady-state real circuit; (c) steady-state imaginary circuit \cite{CMU1}.}
	\vspace{-1em}
\end{figure}

ECF is particularly suitable for SE, as the set of measurements comprises different forms of voltage and current data. Consequently, equivalent circuit models can be derived based on the available voltage-current information representing the measurements and further incorporated within the circuit simulation framework. The circuit equations are then formulated by using some of the common circuit analysis methods, such as Tree-Link Analysis or Modified Nodal Analysis \cite{Pileggi}. The models for transmission lines, transformers and phase shifters, which will be used in the proposed formulation, have been derived in \cite{CMU1}. As a brief insight, the split-circuit model for a transmission line is given in Fig. \ref{fig:1b}-\ref{fig:1c}, where $Y_G=R/(R^2+X^2)$, $Y_B=X/(R^2+X^2)$ and $Y_{sh}=B/2$. 
\subsection{Interval Arithmetic} \label{subsec: Intervals}
Both RTU and PMU measurements contain some degree of uncertainty which has to be taken into consideration in the state estimation process. This error is commonly expressed as standard deviation of a particular measurement most often assuming that the measurement error is Gaussian distributed. Some SE algorithms directly use the available set of measurements, and therefore directly take into account the corresponding uncertainties in the formulations. However, numerous methods imply the transformation of some of the given data, which raises the issue of error propagation calculation.

An alternative way to treat data uncertainty is to limit the range of its possible values. This can be done by representing such data as an interval $[\underline{x},\ \overline{x}]$, where $\underline{x}$ and $\overline{x}$ are the smallest and largest reasonably expected values, respectively. In this way, the data bounds are highlighted and are independent of the actual distribution. If a certain value is a function of such bounded variables and needs to be calculated based on their intervals, it can be done according to the basic rules of interval algebra, as given in \cite{Moore}:
\begin{align} 
    \left[ \underline{x}, \overline{x} \right] + \left[ \underline{y}, \overline{y} \right] &= \left[  \underline{x} + \underline{y},\ \overline{x} + \overline{y} \right] \label{eq:sum_int}\\
    \left[ \underline{x}, \overline{x} \right] - \left[ \underline{y}, \overline{y} \right] &= \left[  \underline{x} - \overline{y},\ \overline{x} + \underline{y} \right] \label{eq:sub_int}\\
    \begin{split} \label{eq:pro_int}
    \left[ \underline{x}, \overline{x} \right] \cdot \left[ \underline{y}, \overline{y} \right] &=  [ \text{min} \left( \underline{x}\cdot\underline{y},\ \underline{x}\cdot\overline{y},\ \overline{x}\cdot\underline{y},\ \overline{x}\cdot\overline{y} \right),\\ &\quad \,\,\, \text{max} {\left( \underline{x}\cdot\underline{y},\ \underline{x}\cdot\overline{y},\ \overline{x}\cdot\underline{y},\ \overline{x}\cdot\overline{y} \right)} ]
    \end{split}\\
    \left[ \underline{x}, \overline{x} \right] / \left[ \underline{y}, \overline{y} \right] &= \left[ \underline{x}, \overline{x} \right] \cdot \left[ \frac{1}{\overline{y}}, \frac{1}{\underline{y}} \right]  \label{eq:div_int}
\end{align}
It should be pointed out that any non-interval value can be interpreted as an interval with equal lower and upper bounds, thus enabling straightforward combination of non-interval values and intervals.

When employing bounds instead of distributions on measurements, naturally the question arises of how to choose these bounds, based on the available information of the measurement mean value $M$ and standard deviation $\sigma$. For a Gaussian probability distribution for example, it can be represented as an interval $[M-3\sigma,\ M+3\sigma]$, since 99.7\% of the normally distributed values are within three standard deviations from the mean \cite{Gaussian}.  

\section{Measurement Modelling} \label{sec: 3}
Present day monitoring systems rely on a diverse set of measurements provided by a variety of different types of monitoring devices at locations spread throughout the system. Modelling the PMU measurements within the equivalent circuit framework can be trivially handled, since the phasor measurements of currents and voltages are consistent with the equivalent circuit state variables. On the other hand, conventional RTU devices usually provide voltage magnitudes, active and reactive power injections, as well as line flows, determined from voltage magnitude and current magnitude measurements and the phase shift between them. This feature makes them suitable for algorithms that are directly based on power flow equations. Hence, the derivation of the appropriate circuit models for RTUs necessitates some transformation of the original data. 
\subsection{PMU Modelling}
\begin{figure}[t!]
    \vspace{-0.175cm}
    \centering
    \begin{subfigure}[h]{0.135\textwidth}
        \centering  
        \includegraphics[width=\textwidth]{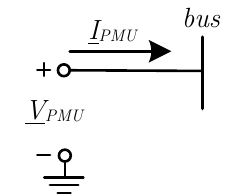}
        \caption{\hspace*{0.25em}}  
        \label{fig:PMU_Inj_a}
    \end{subfigure}
    \hfill
    \begin{subfigure}[h]{0.165\textwidth}  
        \centering 
        \includegraphics[width=\textwidth]{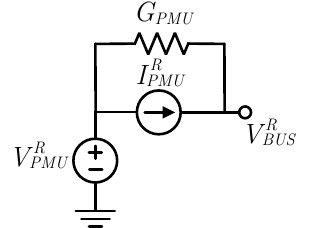}
        \caption{\hspace*{0.25em}}  
        \label{fig:PMU_Inj_b}
    \end{subfigure}
    \hfill
        \begin{subfigure}[h]{0.165\textwidth}  
        \centering 
        \includegraphics[width=\textwidth]{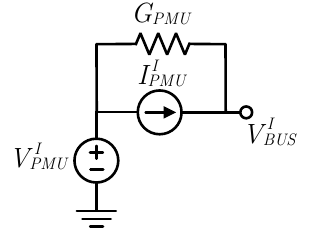}
        \caption{\hspace*{0.25em}}  
        \label{fig:PMU_Inj_c}
    \end{subfigure}
    \caption{PMU injection model: (a) measurement data; (b) real circuit; (c) imaginary circuit.}
    \label{fig:PMU_inj}
\vspace{-0.5em}
\end{figure}

PMUs measure magnitudes and phase angles for both voltages and currents. The measured current can be the current injected or withdrawn at a bus (by a generator or a load) or the current flowing into a transmission line. First, a PMU monitoring only the injection side of the node is considered. In this case, the measurement set comprises bus voltage phasor and injected current phasor, as in Fig.~\ref{fig:PMU_Inj_a}. Since real and imaginary components of the voltage phasor are known, they can be modelled by independent voltage sources. The same applies to the real and imaginary parts of the injection current phasor, which are modelled by independent current sources in real and imaginary split-circuits. In general, a series connection of a voltage and current source renders the voltage source to have no impact. Similarly, it is pointless to connect a current source in parallel with a voltage source. Therefore, a fixed-value conductance $G_{PMU}$ is connected in parallel with the current source, as shown in Fig.~\ref{fig:PMU_Inj_b}\,-\,\ref{fig:PMU_Inj_c}. In this way, both voltage and current information are considered, and the weight of emphasis on each is determined by the value of $G_{PMU}$. These conductances also serve as a kind of a slack element. Currents in these conductances arise when the voltages and currents that are based on possibly inaccurate measurements and imposed in the rest of the circuit are not fully compliant with the physical dependencies between them. For a set of fully accurate measurements, these currents are zero and the voltage $V^a_{BUS}$ is equal to $V^a_{PMU}$, with $a=\{R,I\}$. This will be leveraged later in Sect. \ref{sec: 4} in the formulation of the proposed SE. Note that the resulting circuit model is linear.

Although highly accurate, PMU data still have some small level of inherent inaccuracy. Therefore, instead of fixed values, intervals are used to represent the values of the sources. Based on the mean value and standard deviation of the PMU measurement, interval bounds are determined according to the rule discussed in Sect.~\ref{subsec: Intervals}.

\begin{figure}[t!]
\vspace{-0.3cm}
\centering
\begin{subfigure}[h]{.25\linewidth}
\includegraphics[width=\linewidth]{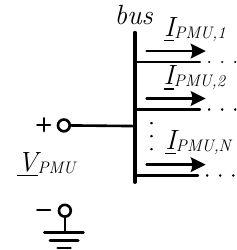}
\caption{\label{fig:PMU_flow_a}}
\end{subfigure}

\begin{subfigure}[h]{.45\linewidth}
\includegraphics[width=\linewidth]{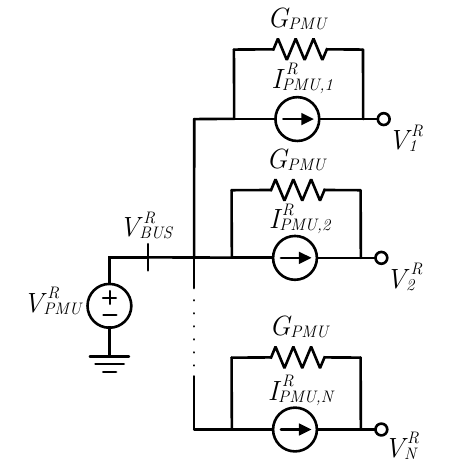}
\caption{\label{fig:PMU_flow_b}}
\end{subfigure}
\begin{subfigure}[h]{.45\linewidth}
\includegraphics[width=\linewidth]{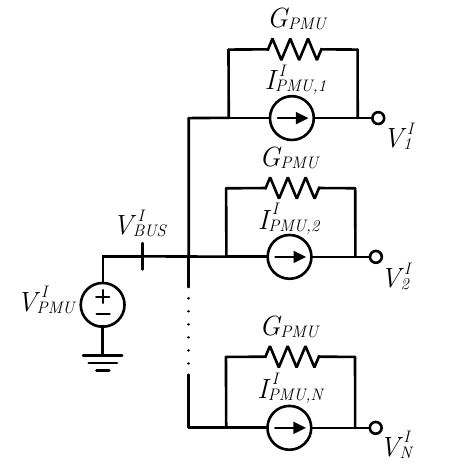}
\caption{\label{fig:PMU_flow_c}}
\end{subfigure}
\caption{PMU line flow model: (a) measurement data; (b) real circuit; (c) imaginary circuit.}
\label{fig:PMU_Inj}
\vspace{-2em}
\end{figure}

The second case is a PMU which measures bus voltage and line current phasors, on all lines incident to the bus, as shown in Fig.~\ref{fig:PMU_flow_a}. This is possible under the assumption that the PMU device has a sufficient number of channels to record the current signals in all lines. Each line flow is modelled by a separate interval-valued current source, with the same fixed-value conductance $G_{PMU}$ added in parallel. The complete model is shown in Fig.~\ref{fig:PMU_flow_b}\,-\,\ref{fig:PMU_flow_c}. The same discussion regarding the currents through the parallel conductances applies as in the previous case.

If a particular PMU contains an insufficient number of channels, the current source is omitted for all the lines missing a current measurement. Instead, the corresponding line terminals are connected directly to the voltage source, thus preserving the effect of the available voltage phasor data. Hence, a variety of potential combinations of measured values can be represented. 

Alternatively, the case with all line flows measured can be interpreted by a PMU injection model with a single aggregated current source, obtained through the summation of all line flow phasor intervals.

\subsection{RTU Modelling}
A set of RTU data comprises bus voltage magnitudes, active and reactive power injections and flows. Therefore, these measurements cannot be directly mapped to an equivalent circuit, which is based on rectangular state variables of voltages and currents. Instead, circuit models will be derived by transforming the available data into a suitable form.

An RTU device essentially measures voltage and current magnitude and the phase shift between voltage and current. The values of active and reactive power are then internally calculated and communicated to the system operator \cite{Caro}. Consequently, power measurements exhibit higher standard deviations than the voltage and current measurements. For the sake of improved accuracy, availability of the originally measured RTU data is assumed in this work.

The case of RTU measurements of the bus voltage magnitude ($V_{RTU}$), injection current magnitude ($I_{RTU}$) and power factor ($\text{cos}\,\phi_{RTU}$) is considered first. Active ($P$) and reactive ($Q$) injection power can be expressed by the following relations:
\begin{align}
    P &= V_{RTU}I_{RTU}\,\text{cos}\left(\phi_{RTU}\right) \label{eq: P_RTU}\\
    Q &= V_{RTU}I_{RTU}\,\text{sin}\left(\phi_{RTU}\right) \label{eq: Q_RTU}
\end{align}

Assuming that the reference direction of the sensed voltage and current corresponds to load conditions, the power drawn at this bus can be expressed as a function of real and imaginary voltage ($V_{R}$ and $V_{I}$, respectively) and real and imaginary current ($I_{R}$ and $I_{I}$, respectively) as follows: 
\begin{align}
    P &= V_{R}I_{R} + V_{I}I_{I} \label{eq: activeP}\\
    Q &= -V_{R}I_{I} + V_{I}I_{R} \label{eq: reacQ}
\end{align}
Equations \eqref{eq: activeP}\,-\,\eqref{eq: reacQ} are further reorganized in order to derive equations for currents in rectangular form as a function of real and imaginary voltages, with $P$ and $Q$ expressed as in \eqref{eq: P_RTU}\,-\,\eqref{eq: Q_RTU}:
\begin{align}
    I_{R} &= \frac{I_{RTU}}{V_{RTU}}\text{cos}\left(\phi_{RTU}\right)V_{R} + \frac{I_{RTU}}{V_{RTU}}\text{sin}\left(\phi_{RTU}\right)V_{I} \label{eq: I_R}\\
    I_{I} &= \frac{I_{RTU}}{V_{RTU}}\text{cos}\left(\phi_{RTU}\right)V_{I} - \frac{I_{RTU}}{V_{RTU}}\text{sin}\left(\phi_{RTU}\right)V_{R} \label{eq: I_I}
\end{align}

\begin{figure}[b!]
    \vspace{-0.175cm}
    \centering
    \begin{subfigure}[h]{0.135\textwidth}
        \centering  
        \includegraphics[width=\textwidth]{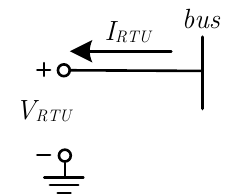}
        \caption{\hspace*{0.25em}}  
        \label{fig:RTU}
    \end{subfigure}
    \hfill
    \begin{subfigure}[h]{0.165\textwidth}  
        \centering 
        \includegraphics[width=\textwidth]{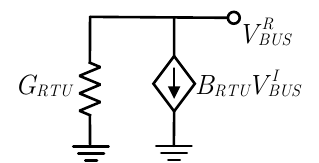}
        \caption{\hspace*{0.25em}}  
        \label{fig:RTU_R}
    \end{subfigure}
    \hfill
        \begin{subfigure}[h]{0.165\textwidth}  
        \centering 
        \includegraphics[width=\textwidth]{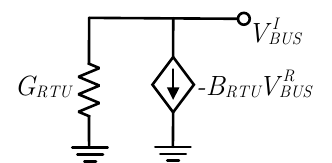}
        \caption{\hspace*{0.25em}}  
        \label{fig:RTU_I}
    \end{subfigure}
    \caption{RTU injection model: (a) measurement data; (b) real circuit; (c) imaginary circuit.}
    \label{fig:RTU_model}
\vspace{-0.5em}
\end{figure}

The first term in equation \eqref{eq: I_R} represents a conductance, since the voltage across it is proportional to the current flowing through it. The second term can be modelled by a dependent current source, governed by the respective voltage in the imaginary circuit, and therefore represents a coupling element between real and imaginary split-circuits. Components of the imaginary circuit can be derived from \eqref{eq: I_I} with the same approach. Finally, the split-circuit model for RTU voltage magnitude and current injection measurements is given in Fig.~\ref{fig:RTU_R}\,-\,\ref{fig:RTU_I}.  

\noindent The values of $G_{RTU}$ and $B_{RTU}$ in these circuits are given by:
\begin{align}
    G_{RTU} &= \frac{I_{RTU}}{V_{RTU}}\text{cos}\left(\phi_{RTU}\right)\label{eq:G_RTU}\\ 
    B_{RTU} &= \frac{I_{RTU}}{V_{RTU}}\text{sin}\left(\phi_{RTU}\right)\label{eq:B_RTU}
    \vspace{-0.5em}
\end{align}
Since the entire derivation has been performed by assuming load reference directions, positive values of $G_{RTU}$ and $B_{RTU}$ indicate consumption of both active and reactive power. Generation results in a negative value for $G_{RTU}$ ($P$ generation) and possibly $B_{RTU}$ ($Q$ generation). Expressions \eqref{eq:G_RTU} and \eqref{eq:B_RTU} are also consistent with modelling the RTU measurement directly with an admittance using the provided voltage and current information.

Similar to the PMU measurements, all measuring errors are modelled by intervals. In that regard, parameters of the RTU sub-circuits are intervals derived from the original measurements, as in \eqref{eq:G_RTU}\,-\,\eqref{eq:B_RTU}, based on the rules of interval arithmetic. 

An RTU device measuring bus voltage and line current flow magnitudes is also a possible scenario. Under the assumption that an RTU monitors all lines incident to its bus, the measured values represented as intervals can be summed up and thus yield an aggregated injection value, which can be mapped into the model given in Fig.~\ref{fig:RTU_model}.

\section{Optimization Problem Formulation} \label{sec: 4}
Models for the provided RTU and PMU measurements are combined with the models for transmission lines, transformers and phase shifters in \cite{CMU1} to construct an equivalent circuit of the observed system. This circuit serves as a starting point in the state estimation procedure. All the elements in the circuit that represent measurements are interval values and therefore, the direct solution of the circuit would result in intervals for the system states. In order to determine the most likely state of the system, we leverage the fact that in a perfect setting the currents through the conductances $G_{PMU}$ should be zero and formulate the following optimization problem:
\begin{subequations}
\label{eq: optimization}
\begin{align}
\min & \sum_{j\in{N^{R,I}_{PMU}}} I^2_{G_{PMU},j} + \sum_{j\in{N_{RTU}}}(G^2_{diff,j} + B^2_{diff,j}) \label{eq: obj_func}
\end{align}
\vspace{-0.5cm}
\begin{alignat}{6}
    &\text{s.t} \quad\quad\quad\quad && I_{ct,lin}(\boldsymbol{X}) = 0 \label{eq: lin_eq}\\
    &&& I_{ct,nlin}(\boldsymbol{X}) = 0 \label{eq: nlin_eq}\\
    &&& x_{min,j}\leq x_j\leq x_{max,j}, &\quad\forall j \in {N_{x}}\label{eq: var_Limit} 
\end{alignat}
\end{subequations}
where $N^{R,I}_{PMU}$ is the number of PMU current sources in both real and imaginary circuits, $N_{RTU}$ is the number of RTU injection sub-circuits in the system and $N_{x}$ is the number of unknown state variables. The vector of equivalent circuit state variables $\boldsymbol{X}$ is equal to:
\begin{equation}
    \boldsymbol{X} = [\boldsymbol{V},\,\boldsymbol{G},\,\boldsymbol{B},\,\boldsymbol{V_{PMU}},\,\boldsymbol{I_{PMU}},\,\boldsymbol{I_{G_{PMU}}},\,\boldsymbol{I_{V_{PMU}}}]^T
\end{equation}
where $\boldsymbol{V}$ is the vector of circuit node voltages and $\boldsymbol{G}$ and $\boldsymbol{B}$ are vectors of RTU conductance and susceptance parameters, respectively; $\boldsymbol{V_{PMU}}$ and $\boldsymbol{I_{PMU}}$ are vectors of PMU voltage and current variables, respectively; $\boldsymbol{I_{G_{PMU}}}$ comprises the currents through all $G_{PMU}$ components, while $\boldsymbol{I_{V_{PMU}}}$ is the vector of currents through the PMU voltage sources.

The quadratic objective function \eqref{eq: obj_func} is convex and consists of two parts. The first term refers to the sum of the currents flowing through the conductances $G_{PMU}$, connected in parallel to the PMU current sources. A general expression for these currents is:
\begin{equation}
    I_{G_{PMU}} = G_{PMU}(V_{PMU,X}-V_{T})
\end{equation}
where $V_{PMU,X}$ is the unknown value of the PMU voltage source connected to the respective current source, while $V_{T}$ is the voltage at the second terminal of the same source. Minimization of all currents flowing through the PMU conductances ensures that for each PMU current source, the difference between voltages at its terminals is as small as possible. Thus, an accurate representation of the underlying PMU measurements is achieved, since the rest of the equivalent circuit representing the system is driven strictly by the PMU voltage and current sources. Since multiple sets of RTU component values, within their respective range, can satisfy the previously explained zero-current condition, the second part of the objective function ensures selection of the solution with the highest probability of occurrence. General expressions for $G_{diff}$ and $B_{diff}$ are:
\begin{align}
    G_{diff} &= G_{X} - G_{M} \label{G_diff}\\
    B_{diff} &= B_{X} - B_{M} \label{B_diff}
\end{align}
where $G_{X}$ and $B_{X}$ are unknown values of the RTU circuit parameters, while $G_{M}$ and $B_{M}$ are the respective most likely values, obtained according to \eqref{eq:G_RTU}\,-\,\eqref{eq:B_RTU} by using the mean values of the RTU measurement intervals. 

Since exact interpretation of the measurement values is a priority, a higher weight must be assigned to the first sum in the objective function. If all PMU sub-circuits contain the same $G_{PMU}$ element, which is assumed here, this term can be shifted in front of the sum and serve as a weight factor. Thus, its value is set accordingly.

The governing circuit equations, obtained by Modified Nodal Analysis, constitute the set of linear \eqref{eq: lin_eq} and nonlinear \eqref{eq: nlin_eq} constraints. The set of KCL equations for the RTU nodes defines \eqref{eq: nlin_eq} with nonlinear terms in the general form $G_{x}V_{x}$ and $B_{x}V_{x}$, where $G_{x}$, $B_{x}$ and $V_{x}$ are the unknown RTU $G$ and $B$  parameters and node voltages, respectively. All remaining circuit equations form the set of linear constraints.

Upper and lower bounds for the variables are defined in \eqref{eq: var_Limit}. The bounds for variables contained in $\boldsymbol{V_{PMU}}$ and $\boldsymbol{I_{PMU}}$ vectors are equal to the PMU measurement limits of the respective voltages and currents, while the range of $\boldsymbol{G}$ and $\boldsymbol{B}$ vector elements are derived based on the RTU measurement intervals, as in \eqref{eq:G_RTU}\,-\,\eqref{eq:B_RTU}. Vectors $\boldsymbol{V}$, $\boldsymbol{I_{G_{PMU}}}$ and $\boldsymbol{I_{V_{PMU}}}$ are not bounded.

\section{Results} \label{sec: 5}
In order to assess the performance of the proposed SE algorithm, the IEEE 14, 57 and 118 bus systems are used, along with the 2869 and 13659 bus systems provided by the Pan European Grid Advanced Simulation and State Estimation (PEGASE) project \cite{13k_buses}.

For the set of time synchronized measurements, it was assumed that current and voltage phasors in rectangular form are available. On the other hand, the available set of conventional measurements consists of bus voltage magnitude, current magnitude and power factor measurements. Each test system comprises both PMU and RTU measurements, as the proposed method requires presence of at least one PMU device in the system. This is due to the fact that at least one voltage source is needed to provide a reference for all other node voltages in the system, and voltage sources are utilized only in the circuit models for PMU measurements. The locations chosen for the measurement devices are listed in Table \ref{tab:measures}. As it can be seen, the number of PMUs is substantially lower than the amount of RTUs, which is a realistic scenario due to the high costs of the time synchronized measurement equipment. It is assumed that each PMU device has a sufficient number of sensing channels to measure currents in all lines incident to its bus. Also, one PMU equipped bus serves as the reference bus for the angles. Apart from the PMU nodes, all other buses are observed by conventional measurements. The percentage of the buses observed by RTU injection current measurements or current flow measurements is given in Table \ref{tab:measures}. Voltage magnitude and power factor values are provided for all RTU buses. Also, for all buses with conventional line current measurements, it is assumed that the current magnitudes in all lines incident to the respective buses are available. Full system observability is maintained in all test cases. 
\begin{table}[t]
\centering
\caption{Measurement Set Composition}
\label{tab:measures}
\begin{tabular}{ |c|c|c|c| }
 \hline
 \multicolumn{1}{|c|}{\multirow{2}{*}{Test Case}} & \multicolumn{1}{c|}{\multirow{2}{*}{\# PMU buses}} & \multicolumn{2}{c|}{\# RTU Buses [\%]}\\
 \cline{3-4}
  &  & Injection & Flow\\ 
 \hline
 14 buses & 3 & 43\,\% & 36\,\%\\ 
 \hline
 57 buses & 7 & 47\,\% & 40\,\%\\ 
 \hline
 118 buses & 10 & 42\,\% & 49\,\%\\ 
 \hline
 2869 buses & 205 & 43\,\% & 50\,\%\\ 
 \hline
 13659 buses & 779 & 44\,\% & 50\,\%\\ 
 \hline
\end{tabular}
\vspace{-0.5em}
\end{table}

\begin{table}[t]
\centering
\caption{Measurement Standard Deviations}
\label{tab:deviations}
\begin{tabular}{|c|c|c|c|c|}
\hline
\multicolumn{3}{|c|}{RTU Measurements} & \multicolumn{2}{c|}{PMU Measurements}\\
\hline
Voltage   & Current   & Power Factor    & Voltage   & Current\\
 \hline
0.4\%   &  0.4\% & 0.6\%   & 0.02\%  & 0.02\%\\
\hline
\end{tabular}
\vspace{-1.5em}
\end{table}
The power flow solution of the test system obtained from MATPOWER served as the basis for creating the measurement sets. All measurements were assumed to have Gaussian distribution with zero mean and a standard deviation that depends on the type of the particular measurement. The adopted values for standard deviations of the different measurements are given in Table \ref{tab:deviations}, which are based on the typical uncertainties of commercially available equipment. One should note the significant difference in accuracy between synchronized and conventional measurements. Finally, the value of the measurement \textit{k} was selected randomly from the range bounded by $\pm\sigma_k$ around the "true" value obtained from the power flow analysis, where $\sigma_k$ is the standard deviation of the measurement \textit{k}.  

Numerous performance indicators are used in literature to evaluate the accuracy of state estimators. In this paper, three different indices are used. The first one is the sum of squared variances of the estimated states:
\begin{equation}
\vspace{-0.7em}
    \sigma^2_{\Sigma} = \sum^{2N}_{i = 1}(\hat{z}_i - z^t_i)^2
\end{equation}
where $\hat{z}_i$ and $z^t_i$ are the estimated and true states, respectively, while $N$ is the number of system buses. Another performance metric is based on the comparison of variances of estimated and measured states:
\begin{equation}
    \xi = \frac{\sum^{M}_{i = 1}(\hat{y}_i - y^t_i)^2}{\sum^{M}_{i = 1}(y^m_i - y^t_i)^2} \label{eq: meas_err}
\end{equation}
where $y^t_i$, $\hat{y}_i$ and $y^m_i$ are true states, estimated states and provided state measurements, respectively, and $M$ is the number of state measurements. Finally, the largest single absolute variance of the estimated states is also observed:
\begin{equation}
    \sigma_{max} = \max_{i\in{2N}}|\hat{z}_i - z^t_i|
\end{equation}
Obviously, all these indices tend to converge to zero as the estimation accuracy is improved. Furthermore, a decrease of the measurement error index given in \eqref{eq: meas_err} indicates the improvement in accuracy provided by the estimated values, compared to the raw measurement data. 

The proposed algorithm was implemented in Matlab. For each test case, the states were estimated in rectangular form. Hence, all bus voltages in polar form can be easily calculated as needed. In order to solve the optimization problem formulated in \eqref{eq: optimization}, the SNOPT solver was used \cite{SNOPT}. To ensure an increased weight factor for the first sum term in \eqref{eq: optimization}, the value of 10 p.u. was selected for the conductance $G_{PMU}$ based on the difference in the order of magnitude of standard deviations of PMU and RTU measurements. Since all PMU measurements have the same standard deviations, assigning the same value to all $G_{PMU}$ elements in the system was justified.

The obtained results are summarized in Table \ref{tab:results}. The order of magnitude of the index $\sigma^2_{\Sigma}$ shows that the accuracy of the proposed method corresponds to those of the methods presented in \cite{Valverde2011}. The values for the $\xi$ factor indicate that the estimated states yield a substantial improvement in the system monitoring accuracy, compared to the case where the available measurements are directly used. Finally, the maximal variances of the estimated states presented in Table \ref{tab:results} verify that even for the state which is estimated with the largest error, the proposed algorithm still yields reasonably small errors. The overall percentage of the PMU buses in the particular test systems used in this paper decreases with the increase of system size, which leads to the continuous increase of values of all three performance indices. These increased values are expected, since a lower number of synchronized measurements results in a less accurate state estimation.  
\begin{table}[t]
\centering
\caption{Averaged Performance Indices}
\label{tab:results}
\begin{tabular}{ |c|c|c|c| }
 \hline
 Test Case & $\sigma^2_{\Sigma}$ & $\xi$ & $|\sigma|_{max}$\\ 
 \hline
 14 buses & 2.0427\,x\,$10^{-6}$ & 0.0285 & 0.0401\,\%\\ 
 \hline
 57 buses & 5.5044\,x\,$10^{-5}$ & 0.1936 & 0.1824\,\%\\ 
 \hline
 118 buses & 9.0264\,x\,$10^{-5}$ & 0.3964 & 0.3136\,\%\\ 
 \hline
 2869 buses & 3.6812\,x\,$10^{-3}$ & 0.6253 & 1.1272\,\%\\ 
 \hline
 13659 buses & 0.0515 & 0.7126 & 1.7380\,\%\\ 
 \hline
\end{tabular}
\vspace{-2em}
\end{table}

\section{Conclusion and Future Work} \label{sec: 6}
This paper introduces a novel formulation for the power system state estimation, based on the equivalent circuit representation of the measurement data. Conventional and synchrophasor measurements are combined into a single-stage estimation problem. The presented numerical results verify a high level of estimation accuracy of the proposed algorithm. 

Future work will include leveraging the reduced nonlinearity of the optimization problem to derive highly efficient and robust solvers of the problem. The formulation also lends itself for improved versions of additional tools usually used in combination with state estimation such as bad data detection, observability analysis, etc. Additionally, detailed models of bus connected devices that provide additional information with regards to voltage and current dependencies can be integrated to further improve the estimation accuracy.

\bibliographystyle{IEEEtran}

\end{document}